\documentclass[a4paper,conference]{IEEEtran}

\IEEEsettopmargin{t}{30mm}
\IEEEquantizetextheight{c}
\IEEEsettextwidth{14mm}{14mm}
\IEEEsetsidemargin{c}{0mm}

\usepackage[utf8]{inputenc}
\usepackage{tabularx}
\usepackage[T1]{fontenc}
\usepackage{url}
\usepackage{hyperref}
\usepackage{ifthen}
\usepackage{cite}
\usepackage{bm} % For bold greek symbols
\usepackage{algorithm}
\usepackage[noend]{algpseudocode}
\usepackage{braket}
\usepackage{xcolor}
\usepackage{graphicx}
\usepackage{multirow}
\usepackage[cmex10]{amsmath} % Use the [cmex10] option 
\newcolumntype{P}[1]{>{\centering\arraybackslash}p{#1}}
\newtheorem{rem}{Remark}
\usepackage{amssymb}
\DeclareMathOperator*{\Mod}{\,mod}
\definecolor{Mycolor1}{HTML}{FFFFFF}
\definecolor{Mycolor2}{HTML}{C41E3A}
\definecolor{Mycolor3}{HTML}{ED9121}
\definecolor{Mycolor4}{HTML}{DFFF00}
\definecolor{Mycolor5}{HTML}{03C03C}
\definecolor{Mycolor6}{HTML}{1560BD}
\definecolor{Mycolor7}{HTML}{9400D3}
\definecolor{Mycolor8}{HTML}{FF1493}
\definecolor{mygreen}{HTML}{248123}
\begin{document}
\title{\fontsize{22.55}{22.55} \selectfont Error Correction for FrodoKEM Using the Gosset Lattice}

\author{\IEEEauthorblockN{Charbel Saliba and Laura Luzzi}
\IEEEauthorblockA{ETIS, UMR 8051, \\ CY Université, ENSEA, CNRS,\\ Cergy, France
 \\
Email: {\{charbel.saliba, laura.luzzi\}@ensea.fr}} \and
\IEEEauthorblockN{Cong Ling}
\IEEEauthorblockA{Department of Electrical \\ and Electronic Engineering \\
Imperial College London, U.K.\\
%London SW7 2AZ, United Kingdom\\
Email: cling@ieee.org}
}

%%% Several authors with up to three affiliations:
% \author{%
%   \IEEEauthorblockN{Perseverance Rover}
%   \IEEEauthorblockA{Mars Technical University\\
%              Space Lab\\
%              Jezero Crater, Mars\\
%              Email: perserverance@splab.mtu.mars}
%   \and
%   \IEEEauthorblockN{Amos Lapidoth and Stefan M.~Moser}
%   \IEEEauthorblockA{ETH Zürich\\
%              ISI (D-ITET), ETH Zentrum\\
%              CH-8092 Zürich, Switzerland\\
%              Email: \{lapidoth, moser\}@isi.ee.ethz.ch}
% }

%%% Many authors with many affiliations:
% \author{%
%   \IEEEauthorblockN{Perserveran Rover\IEEEauthorrefmark{1},
%     Amos Lapidoth\IEEEauthorrefmark{2},
%     Stefan M.~Moser\IEEEauthorrefmark{2}\IEEEauthorrefmark{3},
%     and Donald Duck\IEEEauthorrefmark{4}}
%   \IEEEauthorblockA{\IEEEauthorrefmark{1}%
%              Space Lab\\
%              Jezero Crater, Mars\\
%              Email: perserverance@splab.mtu.mars}
%   \IEEEauthorblockA{\IEEEauthorrefmark{2}%
%              ETH Zürich, 
%              ISI (D-ITET), ETH Zentrum, 
%              CH-8092 Zürich, Switzerland,
%              \{lapidoth, moser\}@isi.ee.ethz.ch}
%   \IEEEauthorblockA{\IEEEauthorrefmark{3}%
%              National Chiao Tung University (NCTU), 
%              Hsinchu, Taiwan, moser@isi.ee.ethz.ch}
%   \IEEEauthorblockA{\IEEEauthorrefmark{4}%
%              Duckburg, Duckingen, Duckland,
%              donald@duck.duck}
% }

\maketitle

\begin{abstract}
We consider FrodoKEM, a lattice-based cryptosystem based on LWE, and propose a new error correction mechanism to improve its performance. Our encoder maps the secret key block-wise into the Gosset lattice $E_8$. 
We propose two sets of parameters for our modified implementation. Thanks to the improved error correction, the first implementation outperforms FrodoKEM in terms of concrete security by $10$ to $13$ bits by increasing the error variance; the second allows to reduce the bandwidth by $7\%$ by halving the modulus $q$. In both cases, the decryption failure probability is improved compared to the original FrodoKEM. 
Unlike some previous works on error correction for lattice-based protocols, we provide a rigorous error probability bound by decomposing the error matrix into blocks with independent error coefficients. 
\end{abstract}

\section{Introduction}

Quantum computers pose a threat since they are capable of breaking most of the cryptographic systems currently in use.
Post-quantum cryptography refers to cryptographic algorithms believed to be secure against a cryptanalytic attack by a quantum computer.
Lattice-based cryptographic constructions are particularly promising candidates for post-quantum cryptography because they offer strong theoretical security guarantees and can be implemented efficiently. 
Therefore, lattice based cryptosystems are considered a safe avenue for replacing the currently used schemes based on RSA and the discrete logarithm.
As of now, NIST is assessing and standardizing PQC algorithms.
In the third round submissions, three of the four finalists in the public-key encryption and key-establishment algorithms are lattice-based schemes, along with the majority of the alternate candidates.

One of the most widely used cryptographic primitives based on lattices is the Learning With Errors problem (LWE), introduced by Regev \cite{LWE}, who proved a worst-case to average-case reduction from the shortest independent vector problem (SIVP) to LWE. It can be used to build a variety of cryptographic algorithms and provides guarantees in terms of IND-CPA and IND-CCA security. 
Later works introduced structured variants of LWE such as Ring-LWE \cite{RLWE} and Module-LWE \cite{module-lwe} which involve ideal lattices and module lattices respectively. Their cryptographic applications are generally more efficient compared to LWE. 
However, in principle the additional algebraic structure might make these variants more vulnerable to attacks. Although currently there are no specific known attacks targeting Ring-LWE or Module-LWE, much progress has been made in recent works to exploit the  structure of ideal lattices and module lattices to solve lattice problems \cite{attack1, attack2, attack3}.
Thus, although the Module-LWE based scheme Kyber \cite{kyber3} was selected as a finalist for the NIST PQC standardization Round 3, the plain-LWE scheme FrodoKEM \cite{frodo3} was selected as an alternate candidate which may provide longer-term security guarantees since it is less susceptible to algebraic attacks. 
From the NIST's perspective, although FrodoKEM can be used in the event that new cryptanalytic results targeting structured lattices emerge, the first priority for standardization is a KEM that would have acceptable performance across widely used applications. This can be done by reducing the communication bandwidth required by the protocol. Moreover, increasing its security level against known attacks would give FrodoKEM a better security margin to resist enhanced computing power 
in the future.

In this paper, we aim at improving the security and/or bandwidth efficiency of FrodoKEM through an enhanced error correction mechanism. A modification of FrodoKEM has been proposed in \cite{Frodo-modification} using Gray labeling and error correcting codes in order to improve the performance. However, the decryption failure analysis in \cite{Frodo-modification} assumes that the coefficients of the error are independent. Unfortunately this assumption does not hold for FrodoKEM, and as shown in \cite{error-dependency}, it can lead to underestimating the decryption failure by a large exponential factor.

In this work, we propose a different approach where enhanced error correction is obtained through lattice encoding and decoding rather than using error-correcting codes. More precisely, our encoder maps the secret key block-wise into the Gosset lattice $E_8$. 
Lattice codes were used in previous works for Ring-LWE based cryptosystems, such as the reconciliation mechanism based on the $\tilde{D}_4$ lattice for NewHope \cite{NEWHOPE}. Due to its optimal density and low-complexity quantization, the $E_8$ lattice was already used in KCL \cite{KCL}, a first round NIST candidate. In our previous work \cite{ourarticle}, the $E_8$ lattice was employed to improve the security of the Module-LWE based candidate KyberKEM. 

The choice of an $8$-dimensional lattice encoder is well-suited to the parameters of FrodoKEM. In fact, due to its particular structure, the error matrix can be decomposed into $8$ blocks of $8$ independent components, which makes a rigorous decryption error analysis possible. The encryption function used by the original FrodoKEM implicitly uses the cubic lattice $\mathbb{Z}^{64} \cong \left( \mathbb{Z}^8 \right)^8$. Accordingly, switching from $\mathbb{Z}^8$ to $E_8$ allows us to improve the security or bandwidth.
We propose two sets of parameters for our modified implementation. Thanks to the improved error correction, the first implementation outperforms FrodoKEM in terms of concrete security by 10-13 bits by increasing the error variance; the second allows to reduce the bandwidth by $7\%$ by halving the modulus $q$.
In particular, we reduce the modulus $q$ from $2^{15}$ to $2^{14}$ for Frodo-640 and from $2^{16}$ to $2^{15}$ for Frodo-976 and Frodo-1344. In both cases, our scheme can ensure a smaller error probability. 
%%%%%%%%%%%%%%%%%%%%%%%%%%%%%%%
%%%%%%%%%%%%%%%%%%%%%%%%%%%%%%%
\subsubsection*{Organization}
This paper is organized as follows. In section \ref{preliminaries}, we provide essential mathematical and cryptographic background for our work, then we develop the proposed modification for FrodoKEM in section \ref{modification-section}. Section \ref{error-section} gives an upper bound  for  the decryption error probability for our algorithm, while section \ref{security_section} derives its security analysis. In the last section, we show the improvements made with regard to security and bandwidth. 
%%%%%%%%%%%%%%%%%%%%%%%%%%%%%%%
%%%%%%%%%%%%%%%%%%%%%%%%%%%%%%%
\section{Notation and preliminaries}  \label{preliminaries}
%This section gives detailed mathematical background that we will use to introduce the FrodoKEM protocol and describe our proposed scheme.

\subsubsection*{Notation} Given a set $A \subseteq \mathbb{R}^n$, $|A|$ stands for its cardinality. All vectors and matrices are denoted in bold. The function sign$(\cdot)$ outputs $1$ for positive real input (including zero) and $-1$ for strictly negative one. For $\mathbf{x} \in \mathbb{R}^n$ we denote $\lfloor \mathbf{x} \rceil$ to be the rounding function of each component of $\mathbf{x}$, where $ \pm 1/2$ is rounded to $0$. 
We also denote $\rfloor \mathbf{x} \lceil$ to be the same as $\lfloor \mathbf{x} \rceil$ except that the worst component of $\mathbf{x}$ - that furthest from an integer - is rounded the wrong way. More formally, if $i_0= \underset{\rm \text{$i$}}{\rm argmax}  \left| x_i - \lfloor x_i \rceil \right|$, then $\rfloor \mathbf{x} \lceil_i= \lfloor x_i \rceil + \text{sign}\left( x_i \right) \cdot \text{sign}\left( |x_i| - \lfloor |x_i| \rceil \right)$ if $i=i_0$ and $\rfloor \mathbf{x} \lceil_i= \lfloor x_i \rceil$ if not.
%$$ \rfloor \mathbf{x} \lceil_i = 
%\begin{cases} 
%\lfloor x_i \rceil &\mbox{if } i \neq i_0 \\
%\lfloor x_i \rceil + \text{sign}\left( x_i \right) \cdot \text{sign}\left( |x_i| - \lfloor |x_i| \rceil \right) & \mbox{if } i=i_0 
%\end{cases}  $$
A constant vector $( \alpha, \dots , \alpha) \in \mathbb{R}^n$ is denoted by $\bm \alpha$. For $a,b \in \mathbb{Z}$, the operation $(a+b) \Mod 2$ is simplified to $a \oplus b$.
\vspace{-3mm}
%%%%%%%%%%%%%%%%%%%%%%%%%%%%%%%
\subsection{Lattice definitions and properties}
An $n$-dimensional lattice $\Lambda$ is a discrete subgroup of $\mathbb{R}^n$ that can be defined as the set of integer linear combinations of $n$ linearly independent vectors, called \emph{basis vectors}. The closest lattice point to $\mathbf{x} \in \mathbb{R}^n$ is denoted by CVP$_{\Lambda}(\mathbf{x})$, and the \emph{Voronoi region} $\mathcal{V}\left( \Lambda \right)$ is the set of all points $\mathbf{x} \in \mathbb{R}^n$ for which CVP$_{\Lambda}(\mathbf{x})=0$.
The \emph{volume} of a lattice, which is a lattice constant, is defined to be the volume of its Voronoi region.
The \emph{Voronoi relevant vectors} of $\Lambda$ are the vectors $\lambda \in \Lambda$ such that $ \langle \mathbf{x}, \lambda \rangle <\| \mathbf{x} \|^2 $ for all $ \mathbf{x} \in \Lambda \setminus \{0, \lambda\}$. 
The minimal distance of the lattice is defined as $\lambda_1( \Lambda ) := \underset{\text{$\mathbf{v} \in \Lambda \setminus \{0\}$}}{\text{min}} || \mathbf{v} ||$.
\vspace{-3mm}
%%%%%%%%%%%%%%%%%%%%%%%%%%%%%%%
\subsection{The Gosset lattice}
We introduce the $8$-dimensional lattice $E_8$ \cite[p.121]{Conway} which will be used throughout this paper. This lattice has a unit volume, and one possible way to generate it is by 
choosing the vector basis as the rows of the matrix 
$$ \mathbf{G}_{E_8}=
\left[\begin{smallmatrix} \vspace{1mm}
2 & 0 & 0 & 0 & 0 & 0 & 0 & 0 \\ \vspace{1mm}
-1 & 1 & 0 & 0 & 0 & 0 & 0 & 0 \\ \vspace{1mm}
0 & -1 & 1 & 0 & 0 & 0 & 0 & 0 \\ \vspace{1mm}
0 & 0 & -1 & 1 & 0 & 0 & 0 & 0 \\ \vspace{1mm}
0 & 0 & 0 & -1 & 1 & 0 & 0 & 0 \\ \vspace{1mm}
0 & 0 & 0 & 0 & -1 & 1 & 0 & 0 \\ \vspace{1mm}
0 & 0 & 0 & 0 & 0 & -1 & 1  & 0 \\ \vspace{1mm}
1/2 & 1/2 & 1/2 & 1/2 & 1/2 & 1/2 & 1/2  & 1/2
\end{smallmatrix}\right]
$$
The Voronoi relevant vectors of $E_8$ form two sets: VR$_1$ which contains the first type of the form $ (\pm 1^2, 0^{6})$, and VR$_2$ which contains $(\pm 0.5^8)$ as the second type. Note that $| \text{VR}_1 |=112$ and $| \text{VR}_2 |=128$, so that the total number of Voronoi relevant vectors is $240$.
\subsection{Error distribution} The error distribution required for the LWE problem defined in the next section is ideally a Gaussian-like distribution. Let $D_{\sigma}(x)=\frac{1}{\sqrt{2 \pi} \sigma}\exp \left( - \|x \|^2 / 2 \sigma^2 \right)$ denotes the probability density function of a zero-mean continuous Gaussian distribution with variance $\sigma$. A rounded Gaussian distribution $\Psi_{\sigma}$ is obtained by rounding a sample from $D_{\sigma}$ to the nearest integer.

As in the FrodoKEM specifications \cite{frodo3}, we use a discrete and symmetric distribution $\chi$ on $\mathbb{Z}$, centered at zero and with finite support $\{-s, \dots ,s\}$, which approximates a rounded Gaussian distribution. In a more detailed manner, we first construct a function $\tilde{\chi}$ on $\{-s, \dots, s\} \subseteq \mathbb{Z}$ from $2^{16}$ samples as follows:
$$ \forall i \in \{-s, \dots, s\}, \; \tilde{\chi}(i) = \frac{1}{2^{16}}\left\lfloor 2^{16} \cdot \int_{\left[ i-\frac{1}{2}, i+ \frac{1}{2} \right]} D_{\sigma}(x) dx \right\rceil.$$
The distribution $\chi$ is obtained from $\tilde{\chi}$ by making small changes in the numerator values of $\tilde{\chi}(i)$ so that the whole sum ends up to be $1$. Note that the chosen distribution is resistant to cache and timing side-channels, while the sampling algorithm for such a distribution is given in \cite[Algorithm 5]{frodo3}. The distance between $ \Psi_{\sigma}$ and $ \chi $ is measured according to the Rényi divergence, which indicates how far a discrete distribution $P$ is from another distribution $Q$ . More formally, for a given positive order $\alpha \neq 1$, the Rényi divergence between $P$ and $Q$ is defined as
$$ D_{\alpha}(P||Q)=\frac{1}{\alpha-1} \ln \left( \sum_{x \in \text{supp} P} P(x) \left( \frac{P(x)}{Q(x)} \right)^{\alpha -1} \right).$$
The Rényi divergence can be used to relate the probabilities of an event according to $P$ or $Q$ \cite[Lemma 5.5]{frodo3}. This justifies why replacing the rounded Gaussian with a distribution which is close in Rényi divergence will preserve the security reductions \cite[Corollary 5.6]{frodo3}.

In our case, we generate $\chi$ for different values of $\sigma$ and compute its Rényi divergence from the rounded Gaussian via the script \texttt{scripts/Renyi.py} in \cite{NEWHOPE}. Note that the support $\{-s, \dots, s\}$ depends on the chosen value of $\sigma$.

\subsection{LWE problem}
The  security  of FrodoKEM and our modified version is based on the hardness of the LWE problem \cite{LWE}. Let $n$ and $q$ be positive integers, and $\chi$ an error distribution over $\mathbb{Z}$.
Take $\mathbf{s}$ to be a uniform vector in $\mathbb{Z}^n_q$.  The problem consists in distinguishing uniform samples $\left( \mathbf{a}, b \right) \leftarrow \mathbb{Z}_q^n \times \mathbb{Z}_q$ from $\left( \mathbf{a}, \langle \mathbf{a}, \mathbf{s} \rangle + e \right)$, where $\mathbf{a}  \xleftarrow{\text{\$}}  \mathbb{Z}_q^n$ is uniform and $e \xleftarrow{\chi} \mathbb{Z}_q$. 
We use a variant of the original LWE problem, for which the secret $\mathbf{s}$ is sampled from $\chi$ rather than $\mathcal{U}$. A polynomial reduction to the original decision LWE is given in \cite{lwe-reduction}.
\subsection{FrodoKEM}
This section presents the basic algorithm of FrodoKEM \cite{frodo3},
%. First of all, it is convenient to mention that FrodoKEM 
which is induced from an IND-CPA secure public-key encryption scheme called FrodoPKE \cite{frodo3} using the Fujisaki-Okamoto (FO) transform \cite{FO_tweaked}, keeping the error probability unchanged. The scheme is designed to guarantee IND-CCA security at three levels: Frodo-640, Frodo-976 and Frodo-1344. The security of these levels matches the brute-force security of AES-128, AES-192 and AES-256 respectively.
Each level is parameterized by an integer dimension $n$ such that $n \equiv 0 \Mod 8$, a variance $\sigma$ and a discrete error distribution $\chi_{\text{Frodo}}$  which is close to the rounded Gaussian $\Psi_{\sigma}$ in Rényi divergence. The LWE modulus $q$ in FrodoKEM is either $2^{14}$ or $2^{15}$ depending on what level is adopted.
A sketch of the algorithm is given in Table \ref{frodoo}. 
Alice generates $\mathbf{A} \xleftarrow{\text{\$}} \mathbb{Z}_q^{n \times n}$, then samples $\mathbf{S},\mathbf{E} \leftarrow \chi_{\text{Frodo}}^{n \times \bar{n}}$, computes the LWE samples $\mathbf{B}=\mathbf{A}\mathbf{S}+\mathbf{E}$ and outputs a public key $(\mathbf{A},\mathbf{B})$. Bob chooses $\mathbf{S}',\mathbf{E}',\mathbf{E}'' \leftarrow \chi_{\text{Frodo}}^{\bar{n} \times n}$, then computes the LWE samples $\mathbf{U}=\mathbf{S}'\mathbf{A}+\mathbf{E'}$ and $\mathbf{V}=\mathbf{S}'\mathbf{B}+\mathbf{E}''$.
A key $\mathbf{k}$ in $\{0,1\}^{\ell}$ is generated unilaterally on Bob's side and encoded into $ \mathbb{Z}_q^{\bar{n} \times \bar{n}} $ using the function \textsc{Frodo.Encode}$(\cdot)$ \cite[Algorithm 1]{frodo3}. Alice recovers  $\mathbf{k'}$ using the decoding mechanism \cite[Algorithm 2]{frodo3}.
The two keys are the same except with probability $P_e=\mathbb{P}\{ \mathbf{k'} \neq \mathbf{k} \}$.
The number of key bits $\ell \in \{ 128, 192, 256 \}$ depends on the assigned security level.
%%%%%%%%%%%%%%%%%%%%%%%%%%%%%%%
\begin{table}[h!] 
\centering
\begin{tabular}{||p{0.179\textwidth}p{0.036\textwidth}p{0.194\textwidth}||} 
\hline 
\multicolumn{3}{||c ||}{Parameters: $q$; $n \in \{640, 976, 1344 \}$; $\bar{n}=8$}\\ [0.5ex] 
\multicolumn{3}{||c ||}{FrodoKEM's distribution $\chi_{\text{Frodo}}$}\\ [0.5ex] 
\hline\hline
\textbf{Alice (server)} &  & \textbf{Bob (Client)} \\ 
$\mathbf{A} \xleftarrow{\text{\$}} \mathbb{Z}_q^{n \times n} $ &  &   \\
$\mathbf{S}, \mathbf{E} \xleftarrow{\text{}} \chi_{\text{Frodo}}^{n \times \bar{n}}$ &  & $\mathbf{S'}, \mathbf{E'} \xleftarrow{\text{}} \chi_{\text{Frodo}}^{\bar{n} \times n}$,  \\
$\mathbf{B} := \mathbf{A} \mathbf{S}+\mathbf{E} \in \mathbb{Z}_q^{n \times \bar{n}}$ &  $\xrightarrow{(\mathbf{A},\mathbf{B})}$ & $\mathbf{E''}  \xleftarrow{\text{}} \chi_{\text{Frodo}}^{\bar{n} \times \bar{n}}$ \\ 
 & & $\mathbf{U} := \mathbf{S'} \mathbf{A}+\mathbf{E'} \in \mathbb{Z}_q^{\bar{n} \times n}$ \\
 & & $\mathbf{V} :=  \mathbf{S'} \mathbf{B} +\mathbf{E''} \in \mathbb{Z}_q^{\bar{n} \times \bar{n}}$\\
 & & $ \mathbf{k} \xleftarrow{\text{\$}} \{0,1\}^{\ell}$ \\
$\mathbf{V'} := \mathbf{C} - \mathbf{U} \mathbf{S} \in \mathbb{Z}_q^{\bar{n} \times \bar{n}}$ & $\xleftarrow{(\mathbf{U},\mathbf{C})}$ & 
 $\mathbf{C}= \mathbf{V}+\textsc{Frodo.Encode}( \mathbf{k})$ \\
 $ \mathbf{k'}= \textsc{Frodo.Decode}(\mathbf{V'})$ & & \\[1ex] 
\hline
\end{tabular}
\vspace{2mm}
\caption{Simplified description of the original FrodoKEM}
\label{frodoo}
\end{table}
\vspace{-4mm}
%%%%%%%%%%%%%%%%%%%%%%%%%%%%%%%
%%%%%%%%%%%%%%%%%%%%%%%%%%%%%%%
\section{Proposed modification of FrodoKEM} \label{modification-section}
With the choice of parameter $\bar{n}=8$ in FrodoKEM \cite{frodo3}, the key $\mathbf{k}\in \{0,1\}^{\ell}$ is encoded into a point of $\mathbb{Z}_q^{64}$. In this section, we propose a modified version of FrodoKEM where the encoder maps the key into a suitably scaled version of the $64$-dimensional lattice $E_8^8$, i.e. the product of $8$ copies of the Gosset lattice. Since $E_8$ is the densest $8$-dimensional packing, this results in a more efficient encoding. Since all integer operations in FrodoKEM are performed modulo $q$, we identify the lattice points that are equivalent modulo $q \mathbb{Z}^{64}$.

Referring to Table \ref{frodoo}, the main adjustments are made for the encryption and decryption algorithms \textsc{Frodo.Encode}$(\cdot)$ and \textsc{Frodo.Decode}$(\cdot)$ respectively. Following the approach in \cite{vanpop}, we search for a suitable scaling parameter $\beta$ such that
$q\mathbb{Z}^{64} \subseteq \left( \beta E_8 \right)^8 \subseteq \mathbb{Z}^{64}$, knowing that $2 \mathbb{Z}^8 \subseteq E_8 \subseteq \frac{1}{2} \mathbb{Z}^8$. 
Our aim is to define an encoding function from $\{0,1\}^{\ell}$ to $\left( \beta E_8 \right)^8/q \mathbb{Z}^{64} \subseteq \mathbb{Z}^{8 \times 8}$. This function is one-to-one if the number of points in $\left( \beta E_8 \right)^8/q \mathbb{Z}^{64}$, which is $\text{Vol}\left(q \mathbb{Z}^{64}\right) / \text{Vol}\left((\beta E_8)^8 \right)$, is greater than or equal to $2^{\ell}$. This condition is verified by setting $\beta= q/2^{\ell / 64} \in \{ q/4, q/8, q/16\}$ for $\ell \in \{ 128, 192, 256\}$. 
The construction of the encoder is as follows. First, $\mathbf{k} \in \{0,1\}^{\ell}$ is partitioned into $8$ substrings $\mathbf{k}_i \in \{0,1\}^{\ell/8}$, $i=0,..,7$. Each substring is mapped into $\beta E_8/q \mathbb{Z}^{8} \subseteq \mathbb{Z}^8$. As an example, for $\ell=128$, the value of $\beta$ is $q/4$. Hence mapping $8$ bits of information into $E_8 / 2\mathbb{Z}^{8}$ allows to map $16$ bits into $E_8 / 4\mathbb{Z}^{8}$.
%, which can be easily extended to map $128$ bits into $\frac{q}{4}E_8 / q\mathbb{Z}^{8}$.
Let $f: \{0,1\}^8 \longrightarrow E_8/2 \mathbb{Z}^8$ that maps $\mathbf{b} =[b_1, b_2, \dots, b_8] \in \{0,1\}^8$ as follows:
\vspace{-1mm}
\[ \begin{cases} 
      f(\mathbf{b}) = [b_1,\dots,b_7,-1] \cdot \mathbf{G}_{E_8} \Mod 2& \text{if } b_1=0 \; \& \& \; b_8=0 \\
      f(\mathbf{b}) = [b_1,\dots,b_7,0] \cdot \mathbf{G}_{E_8} \Mod 2 & \text{if } b_1=0 \; \& \& \; b_8=1 \\
      f(\mathbf{b}) = [b_1,\dots,b_7,1] \cdot \mathbf{G}_{E_8} \Mod 2 & \text{if } b_1=1 \; \& \& \; b_8=0 \\
      f(\mathbf{b}) = [b_1,\dots,b_7,2] \cdot \mathbf{G}_{E_8} \Mod 2 & \text{if } b_1=1 \; \& \& \; b_8=1 
   \end{cases}
\vspace{-1mm}
\]
One can verify that 
%all the $256$ outputs are different. Hence 
$f$ is a bijective function.
%Now since 
%we have the inclusion $E_8 \supseteq 2 \mathbb{Z}^8 \supseteq 4 \mathbb{Z}^8$, one can write:
% $E_8 =  E_8/2 \mathbb{Z}^{8} \oplus 2\mathbb{Z}^8/4\mathbb{Z}^{8} \oplus 4 \mathbb{Z}^8 $, 
 We can map $16$ bits into the quotient $E_8 / 4\mathbb{Z}^{8}$ as follows: map the first 8 bits into 
$ E_8/2 \mathbb{Z}^8$, and the remaining ones into $2 \mathbb{Z}^8 / 4\mathbb{Z}^{8}$. This last mapping is obtained by simply multiplying the input string by $2$.
This example can be extended to the cases $\ell=192$ and $\ell=256$ by considering the chain $E_8 \supseteq 2 \mathbb{Z}^8 \supseteq 4 \mathbb{Z}^8 \supseteq 8 \mathbb{Z}^8 \supseteq 16 \mathbb{Z}^8$. We denote the function that maps the remaining $\ell/8-8$ bits by $g$. The encoding function \textsc{Frodo.Encode}$( \cdot )$ can now be changed to \textsc{E8.Encode}$( \cdot )$ as shown in Algorithm \ref{enc-alg}. 
%Note that our decoding algorithm which will be defined later in Algorithm \ref{dec-alg} requires to operate on an $8 \times 8$ input matrix $\mathbf{N}=\left( N_{i,j} \right)_{0 \leq i, j \leq 7}$ from which it extracts 8 blocks:
%\vspace{-2mm}
%\begin{equation} \label{block}
%  \textsc{Block}_k\left( \mathbf{N} \right)= \left(N_{k \Mod 8,0}, \dots, N_{k+7 \Mod 8,7} \right)  
%  \vspace{-2mm}
%\end{equation}
%where $k=0,\dots ,7$. This forces us to distribute the output matrix of $\textsc{E8.Encode}(\cdot)$ in  in the same way: 
%\vspace{-2mm}
\begin{algorithm} [H] 
\caption{Gosset Lattice Encoding}
\begin{algorithmic}[1] 
\Function{E8.Encode}{$\mathbf{k} \in \{0,1 \}^{\ell}$}     \State $ \mathbf{k}_{i\, : \, i=0, \dots, 7}=(k_{i(\ell/8)}, \dots, k_{i(\ell/8)+\ell /8-1}) \in \{0,1\}^{\ell/8}$
\State $ \mathbf{X}_{i\, : \, i=0, \dots, 7}=f(\mathbf{k}_{i,0}, \dots, \mathbf{k}_{i,7}) \in E_8/2 \mathbb{Z}^8$ 
\State $ \mathbf{X}_{i\, : \, i=0, \dots, 7}'=g(\mathbf{k}_{i,8}, \dots, \mathbf{k}_{i,\ell/8-1}) \in 2\mathbb{Z}^8/2^{\ell/64} \mathbb{Z}^8$ 
\State $\mathbf{R}_{i\, : \, i=0, \dots, 7}=\mathbf{X}_i + \mathbf{X}_i' \in   E_8 / 2^{\ell/64}\mathbb{Z}^8\cong \beta E_8/q\mathbb{Z}^8$ %\Comment{row vectors}
\State \Return $O_{i,j}=\left(R_{(8-i+j) \Mod 8, j}\right)_{\substack{0 \leq i \leq 7 \\ 0 \leq j \leq 7}}$
\EndFunction
\end{algorithmic}
\label{enc-alg}
\end{algorithm}
In the previous algorithm we identify each quotient with the corresponding set of coset leaders in $E_8/2^{\ell/64}\mathbb{Z}^8$.
Note that each substring $\mathbf{k}_i$ is mapped into a vector in $\mathbb{Z}^8$, which is encoded in a block 
$$\textsc{Block}_i\left( \mathbf{O} \right)= \left(O_{i \Mod 8,0}, \dots, O_{i+7 \Mod 8,7} \right)$$ 
of $8$ components of the output matrix $\mathbf{O}$.
Finally, $\textsc{E8.Encode}$ is a bijection from $\{0,1\}^{\ell}$ to $\left( \beta E_8 \right)^8 / q \mathbb{Z}^{64}$.

The decoding algorithm \textsc{E8.Decode} uses the CVP$_{E_8}$ algorithm \cite{Fastcvp} presented in Algorithm \ref{cvp} below.
\begin{algorithm} [H] 
\caption{Closest Vector Point in $E_8$}
\begin{algorithmic}[1]
\Function{CVP$_{E_8}$}{$\mathbf{x} \in \mathbb{R}^{8}$}
\State $\mathbf{f}= \lfloor \mathbf{x} \rceil \, ; \, \mathbf{g}= \rfloor \mathbf{x} \lceil$
\State $\mathbf{y} = \left(1 \oplus \sum f_i \right) \mathbf{f} + \left(1 \oplus \sum g_i \right) \mathbf{g}$
\State $\mathbf{f'}= \lfloor \mathbf{x-\frac{1}{2}} \rceil \, ; \, \mathbf{g'}=  \left \rfloor \mathbf{x-\frac{1}{2}} \right\lceil$
\State $\mathbf{y'} = \left(1 \oplus \sum f'_i \right) \mathbf{f'} + \left(1 \oplus \sum g'_i \right) \mathbf{g'} + \mathbf{\frac{1}{2}}$
\State \Return $\underset{\rm \mathbf{y''} \in \{\mathbf{y}, \mathbf{y'} \}}{\rm argmin} \| \mathbf{x} - \mathbf{y''} \| $
\EndFunction
\end{algorithmic}
\label{cvp}
\end{algorithm}
We describe the decoding protocol in Algorithm \ref{dec-alg}. It concatenates the outputs of $\textsc{CVP}_{E_8}$ to form an element of $\left( \beta E_8 \right)^8 / q \mathbb{Z}^{64}$. Since our lattice $E_8$ is scaled by $\beta$, we use the fact that $\text{CVP}_{\beta E_8}\left( \mathbf{x} \right)= \beta \cdot \text{CVP}_{E_8}\left(\frac{1}{\beta} \mathbf{x} \right)$.
\begin{algorithm} [H] 
\caption{Gosset Lattice Decoding}
\begin{algorithmic}[1]
\Function{E8.Decode}{$\mathbf{N} \in \mathbb{R}_q^{8 \times 8}$}
\State $\mathbf{Y}_{i \, : \, 1 \leq i \leq 8}= \beta \cdot \textsc{CVP}_{E_8}\left( \frac{1}{\beta} \textsc{Block}_i(\mathbf{N})\right) \Mod q$ 
\State $\mathbf{Y}= \left[ \mathbf{Y}_1, \dots ,  \mathbf{Y}_8 \right] \in \left( \beta E_8 \right)^8 / q \mathbb{Z}^{64}$
\State \Return $\mathbf{k'} = \textsc{E8.Encode}^{-1} \left( \mathbf{Y} \right) \in \{ 0,1 \}^{\ell} $
\EndFunction
\end{algorithmic}
\label{dec-alg}
\end{algorithm}
%%%%%%%%%%%%%%%%%%%%%%%%%%%%%%%
%%%%%%%%%%%%%%%%%%%%%%%%%%%%%%%  
  \section{Reliability} \label{error-section}
In this section we aim to provide an upper bound for the decryption error probability for our algorithm. Clearly, an error occurs whenever the received key $\mathbf{k'}$ differs from the original one $\mathbf{k}$, i.e., $P_e=\mathbb{P} \left\{ \mathbf{k} \neq \mathbf{k'} \right\}$. Following Table \ref{frodoo}, the expression of $\mathbf{V}'$ can be simplified as follows: 
\begin{flalign*}
\mathbf{V'} &= \mathbf{C}-\mathbf{U} \mathbf{S} 
= \mathbf{V}+\textsc{E8.Encode}(\mathbf{k}) - \left( \mathbf{S'} \mathbf{A}+\mathbf{E'} \right) \mathbf{S} && \\ \nonumber
&= \mathbf{S'} \left( \mathbf{A} \mathbf{S} + \mathbf{E} \right)\ +\mathbf{E''} + \textsc{E8.Encode}(\mathbf{k}) - \mathbf{S'} \mathbf{A} \mathbf{S} - \mathbf{E'} \mathbf{S} && \\ \nonumber
&= \textsc{E8.Encode}(\mathbf{k}) + \underbrace{\mathbf{S'}\mathbf{E} + \mathbf{E''} - \mathbf{E'} \mathbf{S}}_{\mathbf{E}'''}.
\end{flalign*}
From this we can express the decoded message $\mathbf{k'}$ as
\begin{flalign*}
\mathbf{k'} & =  \textsc{E8.Decode}(\mathbf{V'}) && \\ \nonumber
&=\textsc{E8.Decode}\left( \textsc{E8.Encode}(\mathbf{k}) + \mathbf{E'''} \right) && \\ \nonumber
&= \mathbf{k} + \textsc{E8.Decode}\left(\mathbf{E'''} \right).
\end{flalign*}
Each entry $E'''_{i,j}$ in the matrix $\mathbf{E'''}$ is the sum of $2n$ products of two independent samples from $\chi$, adding to it another independent sample also from $\chi$:
\begin{equation}\label{entries}
 \forall \, 0 \leq i,j \leq 7, \, E'''_{i,j}=\sum_{k=0}^{n-1} \left( S'_{i,k}E_{k,j}-E'_{i,k}S_{k,j}\right) +E''_{i,j}  
\end{equation}
The distribution of $E'''_{i,j}$, denoted by $\chi'$, can be efficiently computed using the product of \emph{probability generating functions}.
Due to equation (\ref{entries}), two entries of the matrix $\mathbf{E'''}$ which are not on the same row or column are independent, and hence we can extract $8$ identically distributed blocks of $8$ independent coordinates from this error matrix, just as indicated in equation (\ref{block}).
Decoding is correct whenever $\textsc{E8.Decode}\left(\mathbf{E'''} \right)=0$. For this it is sufficient to have $\textsc{Block}_k\left( \mathbf{E'''} \right) \in \mathcal{V}\left(\beta E_8\right)$ for all $k=0,..,7$, i.e.,
\vspace{-3mm}
$$\langle \textsc{Block}_k\left( \mathbf{E'''} \right), \mathbf{v} \rangle < \frac{\|\mathbf{v}\|^2_2}{2}, \, \forall \mathbf{v} \in \beta \left( \text{VR}_1 \cup \text{VR}_2 \right).$$
The error probability can thus be bounded by
\vspace{-3mm}
%\begin{equation} \label{pe-bound}
 %   P_e \leq \sum_{i=0}^7 \mathbb{P}\left\{ \Large {\substack{ \exists \, \mathbf{v}_1 \in \text{VR}_1  \, : \, \langle \textsc{Block}_k\left( \mathbf{E'''} \right), \mathbf{v}_1 \rangle \geq \frac{ \beta \|\mathbf{v}_1\|^2_2}{2}
%\\ 
%\text{ or }
%\\
%\exists \, \mathbf{v}_2 \in  \text{VR}_2 \, : \, \langle \textsc{Block}_k\left( \mathbf{E'''} \right), \mathbf{v}_2 \rangle \geq \frac{ \beta \|\mathbf{v}_2\|^2_2}{2}
%} } \right\} 
\begin{small}
\begin{multline} \label{pe-bound}
P_e \leq \sum_{i=0}^7 \mathbb{P}\left\{ \exists \, \mathbf{v}_1 \in \text{VR}_1: \, \langle \textsc{Block}_k\left( \mathbf{E'''} \right), \mathbf{v}_1 \rangle \geq \frac{\beta \|\mathbf{v}_1\|^2_2}{2} \right\} \\ + 
\sum_{i=0}^7 \mathbb{P}\left\{\exists \, \mathbf{v}_2 \in  \text{VR}_2 \, : \, \langle \textsc{Block}_k\left( \mathbf{E'''} \right), \mathbf{v}_2 \rangle \geq \frac{ \beta \|\mathbf{v}_2\|^2_2}{2} \right\}
%\end{equation}
\end{multline}
\end{small}
Since the error probability is independent of the choice of Voronoi relevant vector for vectors of the same type (because the distribution of each entry of $\mathbf{E}'''$ is symmetric, centered at $0$), without loss of generality we can choose $v_1=(1,1,0^6)$ and $v_2=(\scriptsize $1/2$^8)$. This reduces the computations to just two cases.
Choosing the value of $n$ and the modulus $q$, we can compute an upper bound for the above expression for different values of $\sigma$. 
The R.H.S. of equation (\ref{pe-bound}) becomes:
\vspace{-2mm}
\begin{equation*}
\scalebox{0.87}{$8 \cdot 112 \cdot \mathbb{P}\left\{E'''_{0,0}+E'''_{1,1} \geq \beta \right\}   + 8 \cdot 128 \cdot \mathbb{P}\left\{ E'''_{0,0} + \dots + E'''_{7,7} \geq 2 \beta  \right\}.$}
\end{equation*}
%In order to calculate a bound for $ P_e $, one way is to have the exact distribution of $ E_{0,0}''' + E_{1,1}''' $ and $ E_{0,0}''' + \dots + E_{7,7}''' $. The former can be calculated efficiently, while the latter is considerably slow. So what we do is calculate the probability generating function of a sum of four independent entries $ E_{i, j}''' $, which allows us to calculate the probability of the sum of eight independent entries at an exact point using this lemma:
In order to upper bound $P_e$, we use the following.
\begin{rem}
We say that a discrete distribution $p$ taking values in $\mathbb{Z}$ is \emph{unimodal} with mode $0$ if $p(n+1) \leq p(n-1) \; \forall n \geq 0$, and $p(n+1) \geq p(n) \; \forall n<0$. \\
%We say that $p$ is symmetric if $p(n)=p(-n)$ \,$\forall n \in \mathbb{Z}$.
Note that the convolution of two symmetric discrete unimodal distributions is symmetric unimodal \cite[Theorem 4.7]{Unimodality}.
\end{rem}

Since the distribution $\chi'$ is symmetric unimodal, so are the distributions $\chi_2'$, $\chi_4'$, $\chi_8'$ of the sum of two, four and eight independent copies of $E_{i,j}'''$ respectively.
While $\chi_2'$ and $\chi_4'$ can be calculated efficiently, the computation of $\chi_8'$ is slow. Thanks to unimodality, we can estimate the term $\mathbb{P}\left\{ E'''_{0,0} + \dots + E'''_{7,7} \geq 2 \beta  \right\}$ by upper bounding $\chi_8'$ by a piecewise constant function after computing a small number of values. 

%\begin{lemma}
%Let $f(X)=a_0+a_1X+ \dots + a_n X^n$ be a polynomial in $X$ and let $i_0 \in \{0, \dots , 2n\}$ be an even number. The coefficient $\text{Coeff}\left( f^2,X^{i_0} \right) $ of $X^{i_0}$ in $f^2$ is equal to
%$$ \text{Coeff}^2 \left( f,X^{i_0/2} \right) + 2 \sum_{i=0}^{\frac{i_0}{2}-1} \text{Coeff} \left( f,X^{i} \right) \cdot \text{Coeff} \left( f,X^{i_0-i} \right)$$
%\end{lemma}
%At the end, to cover all the points that we need to calculate their probability, we use the fact that the probability generating function is decreasing on the right side in the following meaning:
%\begin{lemma}
%Let $g=\textcolor{red}{a_0}+\textcolor{blue}{a_1} X  + \dots \textcolor{mygreen}{a_{c-1}}X^{c-1} +a_c X^c + \textcolor{mygreen}{a_{c-1}}X^{c+1}+ \dots + \textcolor{blue}{a_1} X^{n-1} + \textcolor{red}{a_0} X^n$. In this case $g$ has symmetric coefficients.
%If $0< a_0 \leq a_1 \leq \dots \leq a_c$, then $g^2$ has also symmetric coefficients with the same ordering, i.e., $\text{Coeff} \left( g^2,X^{0} \right) \leq \text{Coeff} \left( g^2,X^{1} \right) \leq \dots \leq \text{Coeff} \left( g^2,X^{2c} \right)$.
%\end{lemma}
%%%%%%%%%%%%%%%%%%%%%%%%%%%%%%%
%%%%%%%%%%%%%%%%%%%%%%%%%%%%%%% 
\vspace{-3mm}
\section{Security} \label{security_section}
\paragraph{IND-CPA security}
Our scheme only modifies the encoding and decoding functions, the choice of parameters $q$ and $\sigma$, and the error distribution. As shown in \cite{frodo3}, the IND-CPA security of FrodoKEM is upper bounded by the advantage of the decision-LWE problem for the same parameters and error distribution [Theorem 5.9, Theorem 5.10]. We note that the security proof relies on the pseudorandomness of the adversary's observation (similarly to  \cite[Theorem 3.2]{lindner-peikert}) and thus the choice of encoding function has no effect on the  security level, which is only affected by the parameters and error distribution.
%\footnote{In our security analysis, we take a larger standard deviation $\sigma$, and hence we are not worse than before for IND-CPA security.}.
In terms of security against known attacks, the best known bound is given by the BKZ attacks, which involve both primal and dual attacks \cite{primal-dual}.
\paragraph{IND-CCA security}
It was shown in \cite{frodo3} that applying the Fujisaki-Okamoto transformation to the IND-CPA secure protocol FrodoPKE yields an IND-CCA secure key encapsulation mechanism FrodoKEM, even if they use different error distributions, provided that the Rényi divergence between these error distributions is small. In particular, FrodoKEM using the finite support distribution $\chi_{\text{Frodo}}$ is IND-CCA secure provided that the FrodoPKE protocol using a rounded Gaussian distribution $\Psi_{\sigma}$ is IND-CPA secure, and the IND-CCA advantage $\text{Adv}^{\text{ind-cca}}$ can be upper bounded by \cite[Equation (3)]{frodo3} $\forall \alpha>1$:
$$ \small \tfrac{q_{\text{RO}}}{| \mathcal{M}|} + \left( \left( \tfrac{2 \cdot q_{\text{RO}}+1}{| \mathcal{M}|} + q_{\text{RO}} \cdot P_e + 3 \cdot \text{Adv}^{\text{ind-cpa}} \right) \cdot  e^{t \cdot D_{\alpha}(P || Q)}\right)^{1-\frac{1}{\alpha}},$$
where $q_{\text{RO}}$ is the maximum number of oracle queries, $| \mathcal{M}|=2^{\ell}$ is the cardinality of the set of keys, and $t=2n(8 + 8)+64$ is the total number of samples (drawn from the error distribution $\chi$) used to generate $\mathbf{E}, \mathbf{S}, \mathbf{E'}, \mathbf{S'}$ and $\mathbf{E''}$ in Table \ref{frodoo}. In our case, $P=\chi$ and $Q$ is the rounded Gaussian $\Psi_{\sigma}$. The security loss will be minimized by optimizing over the order $\alpha$.
%%%%%%%%%%%%%%%%%%%%%%%%%%%%%%%
%%%%%%%%%%%%%%%%%%%%%%%%%%%%%%%
\section{Performance Comparison}
In this section we show the impact of the of the proposed modification of FrodoKEM in terms of the performance of the protocol. 
We propose two sets of parameters: the first aims at improving the security level and the second at reducing the bandwidth.
Note that for all sets of parameters, $n$ and $\bar{n}$ will remain unchanged. The performance comparison is shown in Table \ref{conclusion_table}. The security level refers to the primal and dual attack via the FrodoKEM script \texttt{pqsec.py} with parameters $n, \sigma$, $q$. 
%%%%%%%%%%%%%%%%%%%%
\paragraph{Parameter set 1 - Improving the security level}
For the first parameter set (see Table \ref{conclusion_table}), we aim at increasing the security level while keeping the same bandwidth and a similar error probability level as in  the original FrodoKEM protocol. To do so, we increase the variance $\sigma$ while keeping $q$ unchanged. Note that we can increase $\sigma$ because of the higher error correction capability provided by our modified encoder.

As shown in Table \ref{conclusion_table}, compared to the original versions of FrodoKEM, the security level is increased by $10$ to $13$ bits, while the error probability is improved.
%%%%%%%%%%%%%%%%%%%%%%%%%%%%%%%
\paragraph{Parameter set 2 - Reducing the bandwidth}
For the second set of parameters in Table \ref{conclusion_table}, we aim at reducing the bandwidth while keeping the same security level. This is achieved by reducing the modulus $q$ by half, which in turn requires a reduction in standard deviation $\sigma$ in order to
preserve a low error probability\footnote{
%The reduction in $\sigma$ is not libre. 
The condition $\sigma \geq 2.12$ is imposed in \cite{frodo3} to allow the reduction from the bounded distance decoding with discrete Gaussian sampling (BDDwDGS) to the decision LWE problem. Note that for efficiency reasons, $\sigma$ is equal to $1.4$ in Frodo-1344, while still guaranteeing a large number $N$ of discrete Gaussian samples, namely $N \approx 2^{111}$. For Frodo-1344 we take $\sigma=1.15$, which still leads to a large number of discrete Gaussian samples, namely $N \approx 2^{75}$.}. Overall, the modulus to noise ratio of the protocol is increased.  
Compared to the original FrodoKEM, this allows to reduce the bandwidth by approximately $7\%$ (see Table \ref{conclusion_table}). 
The communication requirements of the protocol are computed using the functions \texttt{Frodo.Pack} and \texttt{Frodo.Unpack} presented in \cite{frodo3}. In our case we pack both $\mathbf{U} \in \mathbb{Z}^{\bar{n} \times n}$ and $\mathbf{C} \in \mathbb{Z}^{\bar{n} \times \bar{n}}$. Those two vectors, concatenated together, carry about $ \left( \log_2(q) \times n + \log_2(q) \times 8 \right)$ bytes (see \texttt{Frodo.Pack} function).  
As an example, we compute the new bandwidth requirements for the modified Frodo-640: $\log_2(2^{14}) \times 640 + \log_2(2^{14}) \times 8 = 9072$.
%%%%%%%%%%%%%%%%%%%%%%%%%%%%%%%
\begin{table}[h!] \fontsize{8.5}{8.5} \selectfont
    \centering
    \begin{tabular}{|P{0.081\textwidth}|P{0.025\textwidth}|P{0.02\textwidth}|P{0.05\textwidth}|P{0.125\textwidth}|c|}
    \hline
    \multicolumn{6}{|c|}{Original FrodoKEM} \\
    \hline
     & $\sigma$ & $q$  & Security & Bandwidth (bytes) & $P_e$  \\
    \hline
    Frodo-640  & $2.80$ & $2^{15}$ & $145$ & $9720$ & $2^{-138}$ \\
    \hline
    Frodo-976  & $2.30$ & $2^{16}$ & $210$ & $15744$ &  $2^{-199}$ \\
    \hline
    Frodo-1344  & $1.40$ & $2^{16}$ & $275$ & $21632$ &  $2^{-252}$ \\
    \hline
    \multicolumn{6}{|c|}{Security Improvements - Parameter set 1} \\
    \hline
    Modified Frodo-640  & $3.90$ & $2^{15}$ & \textcolor{blue}{$158$} & $9720$ & $2^{-149}$ \\
    \hline
    Modified Frodo-976  & $2.75$ & $2^{16}$ & \textcolor{blue}{$220$} & $15744$ &  $2^{-204}$ \\
    \hline
    Modified Frodo-1344  & $1.68$ & $2^{16}$ & \textcolor{blue}{$287$} & $21632$ &  $2^{-255}$ \\
    \hline
    \multicolumn{6}{|c|}{Reduce Bandwidth - Parameter set 2} \\
    \hline
    Modified Frodo-640  & $2.30$ & \textcolor{blue}{$2^{14}$} & $152$ & \textcolor{blue}{$9072$} & $2^{-152}$ \\
    \hline
    Modified Frodo-976  & $1.80$ & \textcolor{blue}{$2^{15}$} & $215$ & \textcolor{blue}{$14760$} & $2^{-203}$ \\
    \hline
    Modified Frodo-1344  & $1.14$ & \textcolor{blue}{$2^{15}$} & $283$ & \textcolor{blue}{$20280$} &  $2^{-271}$ \\
    \hline
    \end{tabular}
    \caption{Modified parameters for improving the security level and / or bandwidth of FrodoKEM scheme}
    \label{conclusion_table}
\end{table}
%\vspace{-2mm}
\section*{Acknowledgment}
The work of C. Saliba and L. Luzzi is supported by the CY Cergy Paris Université INEX Project AAP 2017.
{
\bibliographystyle{IEEEtran}
\bibliography{References}
}

% trigger a \newpage just before the given reference
% number - used to balance the columns on the last page
% adjust value as needed - may need to be readjusted if
% the document is modified later
%\IEEEtriggeratref{8}
% The "triggered" command can be changed if desired:
%\IEEEtriggercmd{\enlargethispage{-5cm}}

\end{document}